\documentclass[%
,preprint%
 ,
 ,secnumarabic
 ,superscriptaddress%
,amssymb, amsmath,nobibnotes, aps, pre]{revtex4}

\usepackage{epsfig}
\usepackage{psfig}
\usepackage{amsmath}
\usepackage{bm}%
\usepackage{float}

\expandafter\ifx\csname package@font\endcsname\relax\else
 \expandafter\expandafter
 \expandafter\usepackage
 \expandafter\expandafter
 \expandafter{\csname package@font\endcsname}%
\fi

\begin{document}

\psfigdriver{dvips}

\title{Collective Coordinate Control of Density Distributions}%

\author{Obioma U. Uche}%
\affiliation{Department of Chemical Engineering, Princeton University, Princeton, NJ 08544}
\author{Salvatore Torquato}
\affiliation{Department of Chemistry, Princeton University, Princeton, NJ 08544}
\affiliation{Princeton Institute for the Science and Technology of Materials, Princeton University, Princeton, NJ 08544 }
\author{Frank H. Stillinger}
\affiliation{Department of Chemistry, Princeton University, Princeton, NJ 08544}

\date{\today}%

\begin{abstract}
Real collective density variables $C(\boldsymbol{k})$ [c.f. Eq.\ref{Equation3})] in many-particle systems arise from 
non-linear transformations of particle positions, and determine the structure factor $S(\boldsymbol{k})$,
where $\bf k$ denotes the wave vector.
Our objective is to prescribe $C({\boldsymbol k})$ and then to
find many-particle configurations that correspond to such a target $C({\bf k})$
using a numerical optimization technique.
Numerical results reported here extend earlier one- and two-dimensional studies to include three dimensions.  In 
addition, they demonstrate the capacity to control $S(\boldsymbol{k})$ in the neighborhood of $|\boldsymbol{k}| =$ 
0.  The optimization method employed generates multi-particle configurations for which $S(\boldsymbol{k}) \propto 
|\boldsymbol{k}|^{\alpha}$, $|\boldsymbol{k}| \leq K$, and $\alpha =$ 1, 2, 4, 6, 8, and 10.  The case $\alpha =$ 1 is 
relevant for the Harrison-Zeldovich model of the early universe, for superfluid $^{4}\mbox{He}$, and for jammed amorphous 
sphere packings.  The analysis also provides specific examples of interaction potentials whose 
classical ground state are configurationally degenerate  and disordered.
\end{abstract}

\maketitle

{\renewcommand{\baselinestretch}{2} \small \normalsize}

\renewcommand{\thesection}{\arabic{section}}
\section{\label{sec:Section1}Introduction}
\hspace{17.5pt}

\numberwithin{equation}{section}

\par{ Spatial arrangements of particles in many-body systems exhibit wide diversity that arises from the 
interactions that are present, and from the prior history of those systems.  One of the available analytical 
tools that has proved useful for describing those spatial arrangements, whether for individual cases or for 
ensemble averages, is the set of collective density variables.  These are conventionally defined for $N$ 
identical particles in the following way:

\begin{equation}
   \rho(\boldsymbol{k}) = \sum_{j=1}^{N} \exp (i \boldsymbol{k} \cdot \boldsymbol{r}_{j})
\label{Equation1}
\end{equation}

\noindent
where $\boldsymbol{r}_j$ denotes the location of particle $j$, and the $\boldsymbol{k}$ are the wave vectors 
appropriate for the containing volume and boundary conditions.}

\par{ Collective density variables have played an important role in a variety of problems in condensed matter 
physics.  Specifically, they have been used in illustrating that large-scale density variations in superfluid 
$^{4}\mbox{He}$ are in fact long-wavelength phonons~\cite{Fe72}.  Introduction of the appropriate collective coordinates 
is natural in describing independent plasma oscillations brought about by the long-range Coulomb interactions 
between electrons in metals~\cite{Pi55}.  Furthermore, application of these variables has aided in the derivation 
of self-consistent integral equations for pair correlation functions in classical fluids~\cite{Pe58}, in obtaining 
corrections to the random phase approximation for the electron gas~\cite{An70}, and in generating classical 
ground states for particle systems~\cite{Fa91,Uc04,Su05}.  But in spite of the fact that collective density variables 
appear widely in the literature~\cite{Fe72,Pi55,Pe58,An70,Fa91,Uc04,Su05,Ma76,Le83,De96}, the nonlinearity of the 
transformation in Eq.~(\ref{Equation1}) from particle positions to collective variables entails nontrivial 
mathematical properties that are still incompletely understood~\cite{Fa91,Uc04}.  The present paper is devoted 
to clarifying some of the remaining issues.}

\par{ Many physical applications of collective density variables focus on the structure factor $S(\boldsymbol{k})$ 
for the many-particle system involved:

\begin{equation}
\begin{split}
   S(\boldsymbol{k}) &= \frac{|\rho(\boldsymbol{k})|^2}{N}     \\
                     &= 1 + \frac{2}{N} C(\boldsymbol{k})      \\
\end{split}                     
\label{Equation2}
\end{equation}

\noindent
where the real quantities $C(\boldsymbol{k})$ are the following:

\begin{equation}
  C(\boldsymbol{k}) = \sum_{j=1}^{N-1} \sum_{l=j+1}^{N} \cos[\boldsymbol{k} \cdot (\boldsymbol{r}_{j} - \boldsymbol{r}_{l})].
\label{Equation3}
\end{equation} 

\noindent
In view of the fact that the phase angles of the $\rho(\boldsymbol{k})$ are irrelevant for most applications, 
it suffices to focus attention on the $C(\boldsymbol{k})$'s.}

\par{ A considerable challenge involves determining what sets of $C(\boldsymbol{k})$ 
values correspond to attainable particle configurations, and how to generate and 
describe those special configurations, including ground-state
structures. In particular, 
it is important to understand the extent to which these real collective variables at small wave vectors 
$\boldsymbol{k}$ are controllable.  Although we begin by considering the general situation, much of the attention 
in the following will involve the examination of ``hyperuniform'' systems~\cite{To03,Fn01}, namely those in 
the infinite system limit for which:

\begin{equation}
   \lim_{\boldsymbol{k} \rightarrow 0} S(\boldsymbol{k}) = 0.
\label{Equation4}
\end{equation}

\noindent
This defining characteristic of hyperuniformity states that the usual mean-square particle-number fluctuations 
increases less than  $R^{d}$, where $R$ denotes the linear
size of an observation window and $d$ is the space dimension~\cite{To03}.  Considering the fact that various 
hyperuniform physical systems exhibit characteristic $\boldsymbol{k}$ dependence of their structure factors near 
the origin ({\it e.g.}, the ground state of liquid $^4\mbox{He}$~\cite{Fe54,Fe56,Re67} 
as well as random, jammed hard-sphere  packings~\cite{Do05} and the early Universe~\cite{Ga02}), 
it becomes important to understand what 
$N$-particle configurational implications stem from these specific $S(\boldsymbol{k})$ forms.  
This has guided the analysis detailed below.}

 Because our objective is to prescribe $C({\boldsymbol k})$ [or, equivalently $S({\boldsymbol k})$]
and then to find many-particle configurations that may correspond to such a target
structure function, this problem can be regarded to be an {\it inverse problem}. 
The analogous inverse problem in real space in which the pair correlation function
is prescribed has received considerable attention in the last several years 
\cite{TB02,To02,Cr03,Uc06,To06}. This class of inverse problems has come
to be called {\it construction} problems \cite{TB02,To02,Cr03,Uc06}. {\it A priori},
a prescribed pair structural function is not necessarily realizable by a many-particle
configuration. A solution to the construction problem therefore provides 
numerical evidence for realizabilty of a target pair structural function. Thus,
the present investigation has important implications for the {\it realizability} problem
of statistical mechanics, which seeks to determine the necessary conditions
that a prescribed pair correlation [or its equivalent Fourier representation, $S({\bf k})$]
must possess in order for it to correspond to a many-particle system \cite{To02,To06,Co04}.

The present paper extends our earlier one-dimenisonal \cite{Fa91} and two-dimensional
\cite{Uc04} studies of collective coordinate control of density distributions.
Our focus in these two studies was to consider continuous
and bounded pair potentials and to constrain the corresponding collective parameters 
$C(\boldsymbol{k})$, with wave vector $\boldsymbol{k}$
magnitudes at or below a chosen cutoff, to their absolute minimum values. In other
words, density fluctuations for those
$\boldsymbol{k}$'s were completely suppressed. In our two-dimensional
investigation \cite{Uc04}, we were able to distinguish between three different
ground-state structural regimes as the number of constrained 
wave vectors were increased - disordered, wavy crystalline, and
crystalline regimes. Evidence for a disordered or irregular ground state,
a counterintuitive notion, had heretofore not been provided. In the present
work, we extend these results not only to three-dimensional
ground state problems but to more general two- and three-dimensional 
hyperuniform many-particle systems.

\par{ The next section provides some of the mathematical structure needed to understand how fixing 
values of sets of the real collective variables $C(\boldsymbol{k})$ exerts control over the allowed many-particle 
configurations.  A description of our numerical methods for analyzing this problem follows in Section~\ref{sec:Section3}.  
Results of the numerical study cover both two- and three-dimensional systems, and are presented respectively in 
Sections~\ref{sec:Section4} and~\ref{sec:Section5}.  
Among other results,  we provide specific examples of interaction potentials whose
classical ground state are configurationally degenerate and disordered.
Our conclusions and discussion of some remaining issues appear 
in Section~\ref{sec:Section6}.}

\renewcommand{\thesection}{\arabic{section}}
\section{\label{sec:Section2}General Relations}
\hspace{17.5pt}

\numberwithin{equation}{section}

\par{ Suppose that the $N$ point particles reside in a one-, two-, or three-dimensional container that is an $L_x$ interval, 
an $L_x \times  L_y$ rectangle, or an $L_x \times L_y \times L_z$ rectangular solid.  Furthermore, suppose that periodic 
boundary conditions apply.  The applicable wave vectors have components:

\begin{equation}
   k_\gamma = \frac{2 \pi n_\gamma}{L_\gamma}    \hspace{10pt}  (n_\gamma = 0,\pm{1},\pm{2},\cdots)
\label{Equation5}
\end{equation}   

\noindent     
where $\gamma$ = $x$, $y$, $z$ as needed.  It is easy to see that the $C(\boldsymbol{k})$ must obey the following 
properties:  

\begin{equation}
 \begin{gathered}
    C(0) = \frac{1}{2} N (N-1) \hspace{5pt}, \\
    C(\boldsymbol{k}) = C(-\boldsymbol{k})  \hspace{5pt}.
 \end{gathered}
\label{Equation6}
\end{equation}

\noindent
Furthermore, these collective variables are necessarily confined to the range:

\begin{equation}
   -\frac{1}{2} N \leq C(\boldsymbol{k}) \leq \frac{1}{2} N (N-1) \hspace{20pt} (\boldsymbol{k} \neq 0)
    \hspace{5pt},
\label{Equation6a}
\end{equation}

\noindent
and as they vary over this range they measure the magnitude of density inhomogeneity at wave vector 
$\boldsymbol{k}$ in the $N$-particle system.}

\par{ Although the number of collective variables is infinite, the $N$-particle system possesses only $dN$ 
configurational degrees of freedom, where $d$ is the Euclidean space dimension ($d$ = 1,2,3).  Consequently, 
it is unreasonable to suppose (barring special circumstances) that generally all $C(\boldsymbol{k})$'s could be 
independently controlled.  However it is possible, as examples in Refs.~\cite{Fa91,Uc04} and in subsequent sections 
below will illustrate, to specify simultaneously a number of the collective variables equal to a significant 
fraction of $dN$.  In particular, let $\textbf{Q}$ be a finite set of the $\boldsymbol{k}$'s meeting this 
criterion, and let $C_{0}(\boldsymbol{k})$ be the target value to which $C(\boldsymbol{k})$ is to be constrained.  
Of course, each $C_{0}(\boldsymbol{k})$ must lie in the range specified by inequalities~(\ref{Equation6a}) above.  
Then consider the following non-negative objective function: 

\begin{equation}
 \begin{gathered}
   \Phi(\boldsymbol{r}_1 \ldots \boldsymbol{r}_N) = \sum_{\boldsymbol{k} \in \textbf{Q}} 
   V(\boldsymbol{k}) [C(\boldsymbol{k}) - C_{0}(\boldsymbol{k})]^{2} \hspace{5pt}, \\
   V(\boldsymbol{k}) = V(-\boldsymbol{k}) > 0 \hspace{5pt}.
 \end{gathered}   
\label{Equation7}
\end{equation}    

\noindent
This continuous and differentiable function of the particle coordinates attains its absolute minimum value zero if and 
only if the $C(\boldsymbol{k})$ for all $\boldsymbol{k} \in \textbf{Q}$ equal their target values.}

\par{ By inserting the definitions~(\ref{Equation3}) for the collective variables into Eq.~(\ref{Equation7}), one trivially 
has the following

\begin{equation}
   \Phi(\boldsymbol{r}_1 \ldots \boldsymbol{r}_N) = \sum_{\boldsymbol{k} \in \textbf{Q}} 
   V(\boldsymbol{k}) \{ \sum_{j<l}^{N} \sum_{m<n}^{N} \cos[\boldsymbol{k} \cdot (\boldsymbol{r}_j - \boldsymbol{r}_l)]
   \cos[\boldsymbol{k} \cdot (\boldsymbol{r}_m - \boldsymbol{r}_n)] - 2 C_{0}(\boldsymbol{k}) 
   \sum_{j<l}^{N} \cos[\boldsymbol{k} \cdot (\boldsymbol{r}_j - \boldsymbol{r}_l] + C_{0}^{2}(\boldsymbol{k})\} \hspace{5pt}.
\label{Equation8}
\end{equation}

\noindent
The right member of this last equation can be resolved into symmetric combinations of four, three, and two particle 
contributions, plus an additive constant:   

\begin{equation}
   \Phi(\boldsymbol{r}_1 \ldots \boldsymbol{r}_N) = \sum_{j<l<m<n}^{N} v_{4}(\boldsymbol{r}_j,\boldsymbol{r}_l,
   \boldsymbol{r}_m,\boldsymbol{r}_n) + \sum_{j<l<m}^{N} v_{3} (\boldsymbol{r}_j,\boldsymbol{r}_l,\boldsymbol{r}_m) 
   + \sum_{j<l}^{N} v_{2} (\boldsymbol{r}_j,\boldsymbol{r}_l) + \Phi_{0} \hspace{5pt}.
\label{Equation9}
\end{equation}

\noindent
The specific forms of these contributions are as follows:

\begin{equation}
\begin{aligned}
     v_{4}(\boldsymbol{r}_j,\boldsymbol{r}_l,\boldsymbol{r}_m,\boldsymbol{r}_n) 
     &= 2 \sum_{\boldsymbol{k} \in \textbf{Q}} V(\boldsymbol{k}) \{ \cos[\boldsymbol{k} \cdot (\boldsymbol{r}_j - \boldsymbol{r}_l)] \cos[\boldsymbol{k} \cdot (\boldsymbol{r}_m - \boldsymbol{r}_n)] \\
     & \quad + \cos[\boldsymbol{k} \cdot (\boldsymbol{r}_j - \boldsymbol{r}_m)] \cos[\boldsymbol{k} \cdot (\boldsymbol{r}_l - \boldsymbol{r}_n)] 
     + \cos[\boldsymbol{k} \cdot (\boldsymbol{r}_j - \boldsymbol{r}_n)] \cos[\boldsymbol{k} \cdot (\boldsymbol{r}_l - \boldsymbol{r}_m)] \}   \hspace{5pt}, \\
     v_{3}(\boldsymbol{r}_j,\boldsymbol{r}_l,\boldsymbol{r}_m) 
     &=  2 \sum_{\boldsymbol{k} \in \textbf{Q}} V(\boldsymbol{k}) \{ \cos[\boldsymbol{k} \cdot (\boldsymbol{r}_j - \boldsymbol{r}_l)] \cos[\boldsymbol{k} \cdot (\boldsymbol{r}_j - \boldsymbol{r}_m)] \\
     & \quad + \cos[\boldsymbol{k} \cdot (\boldsymbol{r}_l - \boldsymbol{r}_j)] \cos[\boldsymbol{k} \cdot (\boldsymbol{r}_l - \boldsymbol{r}_m)] 
     + \cos[\boldsymbol{k} \cdot (\boldsymbol{r}_m - \boldsymbol{r}_j)] \cos[\boldsymbol{k} \cdot (\boldsymbol{r}_m - \boldsymbol{r}_l)] \}   \hspace{5pt}, \\
     v_{2}(\boldsymbol{r}_j,\boldsymbol{r}_l) &=  \sum_{\boldsymbol{k} \in \textbf{Q}} 
     V(\boldsymbol{k}) \{ \cos ^{2}[\boldsymbol{k} \cdot (\boldsymbol{r}_j - \boldsymbol{r}_l)] 
     - 2 C_{0} \cos[\boldsymbol{k} \cdot (\boldsymbol{r}_j - \boldsymbol{r}_l)] \}  \hspace{5pt}, \\
     \Phi_{0} &= \sum_{\boldsymbol{k} \in \textbf{Q}} V(\boldsymbol{k}) C_{0}^{2}(\boldsymbol{k}) \hspace{5pt}. \\
\end{aligned}
\label{Equation10}
\end{equation}
    
\noindent
If $\Phi$ is interpreted as a potential energy of interaction for the $N$ point particles, then Eqs.~(\ref{Equation9}) 
and~(\ref{Equation10}) specify four, three, and two particle interaction potentials operating in the system.  If 
classical ground-state configurations for the $N$ particles subject to that potential exist for which $\Phi$ = 0, 
then those configurations necessarily attain the desired target values of the collective variables.}

\par{ The remainder of our analysis will be restricted to cases for which the wave vector set $\textbf{Q}$ consists of 
all those $\boldsymbol{k}$ lying within a given distance from the origin:

\begin{equation}
   \textbf{Q} : 0 \leq |\boldsymbol{k}| \leq K \hspace{5pt}.
\label{Equation11}
\end{equation}   

\noindent
In addition, we shall also confine attention to the following specific family of forms for $C_{0}(\boldsymbol{k})$:

\begin{equation}
   C_{0}(\boldsymbol{k}) = -N/2 + D |\boldsymbol{k}|^{\alpha} \hspace{5pt},
\label{Equation12}
\end{equation}

\noindent
where the multiplicative constant $D$ and the exponent $\alpha$ are both non-negative.  This choice focuses on the 
behavior of the particle system in the hyperuniform regime.}

\par{ Previous work~\cite{Fa91,Uc04} only considered the situation where $D$ = 0, $i.e.$, each of the constrained 
collective variables was required to be at its minimum value $-N/2$.  In that event, it is possible to utilize a 
simpler potential energy (objective) function $\tilde{\Phi}$ that reduces to a sum of just two-particle interactions: 

\begin{equation}
   \tilde{\Phi}(\boldsymbol{r}_1 \ldots \boldsymbol{r}_N) = \Omega^{-1} \sum_{\boldsymbol{k} \in \textbf{Q}} 
   \tilde{V}(\boldsymbol{k}) C(\boldsymbol{k}) \hspace{5pt}, 
\label{Equation13}  
\end{equation}

\noindent
In this expression $\Omega$ is the system length ($L_x$), area ($L_x L_y$), or volume ($L_x L_y L_z$), and 

\begin{equation}
   \tilde{V}(\boldsymbol{k}) = \tilde{V}(-\boldsymbol{k}) > 0 \hspace{5pt}.
\label{Equation14}
\end{equation} 

\noindent
It is easy to show~\cite{Fa91,Uc04} that Eq.~(\ref{Equation13}) is equivalent to: 

\begin{equation}
 \begin{gathered}
   \tilde{\Phi}(\boldsymbol{r}_1 \ldots \boldsymbol{r}_N) = \sum_{j<l}^{N} 
   \tilde{v}_{2} (\boldsymbol{r}_j, \boldsymbol{r}_l) \hspace{5pt}, \\
   \tilde{v}_{2}(\boldsymbol{r}_j, \boldsymbol{r}_l) = \Omega^{-1} \sum_{\boldsymbol{k} \in \textbf{Q}} 
   \tilde{V}(\boldsymbol{k}) \exp[i \boldsymbol{k} \cdot (\boldsymbol{r}_j - \boldsymbol{r}_l)] \hspace{5pt}.
 \end{gathered}  
\label{Equation15}
\end{equation}

\noindent
On account of the positivity condition~(\ref{Equation14}) on $\tilde{V}(\boldsymbol{k})$, the absolute minimum of 
$\tilde{\Phi}$ will have the value:

\begin{equation}
   \min_{\boldsymbol{r}_1 \ldots \boldsymbol{r}_N}(\tilde{\Phi}) = -N/2 \sum_{\boldsymbol{k} \in \textbf{Q}}
   \tilde{V}(\boldsymbol{k})
\label{Equation16}
\end{equation}

\noindent
if and only if there exist particle configurations satisfying all of the collective variable constraints.  Of course 
the more elaborate (and more general) formulation defined by Eqs.~(\ref{Equation7})-(\ref{Equation10}) is also 
applicable to this $D$ = 0 case.}

\par{ Suppose that system configurations have been found which succeed in producing the desired target values for 
collective variables over the wave vector set $\textbf{Q}$.  Then these configurations obviously satisfy the same 
target values over the smaller wave vector set $\textbf{Q}^{\prime}$ $\subset$ $\textbf{Q}$.  But in view of the 
fact that $\textbf{Q}^{\prime}$ entails fewer configurational constraints, one can expect that a more inclusive set of 
$N$-particle configurations satisfies those constraints.  Indeed that is exactly what has been found in our previous 
one- and two-dimensional studies~\cite{Fa91,Uc04}, and further two- and three-dimensional examples will be reported in 
this paper.  With respect to the objective functions $\Phi$ and $\tilde{\Phi}$ whose minimization indicates a solution 
to the constraint problem, the corresponding potential energy interpretations demonstrate the presence of 
configurationally $degenerate$ classical ground states for the $N$ particles,
including disordered or highly irregular ground-state structures.
This ground-state degeneracy phenomenon 
has been the subject of a recent theoretical 
study~\cite{Su05} that proceeds in a rather different direction than from the 
collective coordinate perspective presented here. The key idea used by
S\"{u}t\'{o} to prove a theorem about ground states is that the Fourier transform of the pair
potential V(k) be nonnegative with compact support, which
was first employed in Ref. \cite{Fa91}; see also Ref. \cite{Uc04}.}

\renewcommand{\thesection}{\arabic{section}}
\section{\label{sec:Section3}Computational Methods}
\hspace{17.5pt}

\numberwithin{equation}{section}

\par{ In all calculations reported below, we assume that the system region $\Omega$ is constrained to a unit square 
in two dimensions or a unit cube in three dimensions, to which periodic boundary conditions are applied.  Special 
attention is devoted to the choice of the system population $N$ for both two and three dimensions.  In two dimensions, 
$N$ has been chosen such that all particles could be arranged in the square simulation box in an aligned and nearly 
undeformed version of the triangular lattice.  For this purpose, the system occupancy $N$ is required to be the 
product of integers $2pq$ where the rational number $p/q$ is a close approximation to the irrational number $3^{1/2}$.  
The corresponding configuration consists of $2q$ rows of $p$ particles.  The same approach 
was used in selecting the system size in our earlier two-dimensional study~\cite{Uc04}.  In three 
dimensions, $N$ is selected such that all particles can be assembled as an aligned face-centered cubic lattice.  Thus, 
the system occupancy is $N = 4s^{3}$ particles where $s$ is a non-zero integer.}

\par{ For any given choice of the independent parameters $N$, $K$, $D$, and $\alpha$, the majority of our numerical 
studies utilized a random number generator to create an initial configuration of the particles inside the simulation 
space $\Omega$.  As one would expect, this starting point typically produces a large positive value of the objective 
function of interest.  We consider the specific case of Eq. (\ref{Equation7}) where $V(k)$ is unity for the 
set $Q$ under consideration.  Introduction of the original configuration to the numerical optimization tool 
MINOP~\cite{De79, Ka99} results in a search for a particle pattern at the absolute minimum of the objective function.  
Our earlier two-dimensional study~\cite{Uc04} involved use of the conjugate gradient method~\cite{Pr86} as our 
numerical tool of choice.  The greater utility of the MINOP optimization technique for the present investigation has 
been an important advantage that merits brief discussion.}  

\begin{figure}[H]
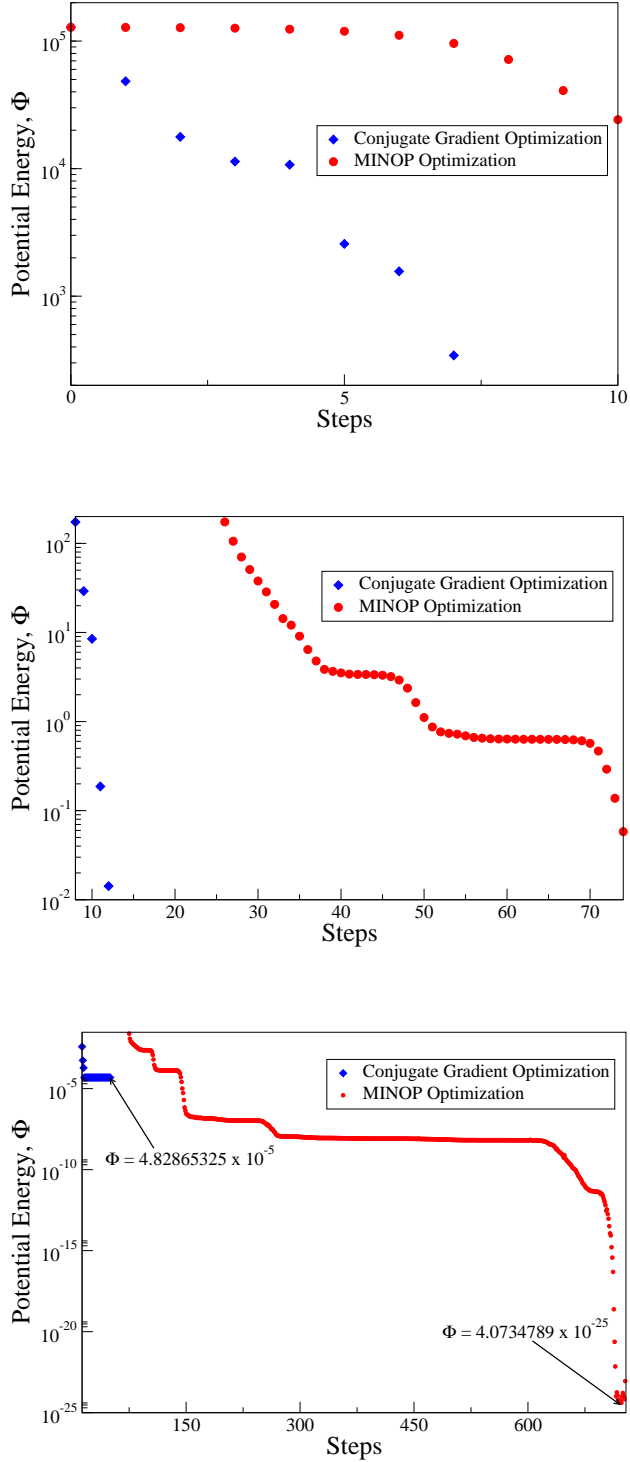

\begin{center}
\includegraphics[width=.5\linewidth]{PTrack1.eps}
\end{center}
\vspace{0.05in}
\begin{center}
\includegraphics[width=.5\linewidth]{PTrack2.eps}
\end{center}
\vspace{0.05in}
\begin{center}
\includegraphics[width=.5\linewidth]{PTrack3.eps}
\end{center}
\label{fig:PTrack123}
\caption{ (Color online) Tracking of the potential energy for conjugate gradient and MINOP algorithms.  Top panel: Early steps.  
Middle panel: Intermediate steps.  Bottom panel: Late Steps.  It should be noted that both algorithms were applied 
to the same $d = 1$ minimization problem and initial particle configuration. The relevant parameters for the minimization 
problem are $N$ = 200, $K$ = 10$\pi$, $\alpha$ = 6, and $D K^6=$ 0.01.}
\end{figure}

\par{ The three plots in Fig.~1 display the tracked potential energy $\Phi$ during the course of a minimization 
for both conjugate gradient (CG) and MINOP optimizations.  Here, we have applied the algorithms to $\Phi$ minimization 
for a one-dimensional system with the parameters $N$ = 200, $K$ = 10$\pi$, $\alpha$ = 6, and $D K^6=$ 0.01.  The 
final values of the objective function $\Phi$ are typical for several cases examined and clearly indicate that the 
MINOP strategy is better suited to finding a numerically precise solution to the problem than is the conjugate gradient 
approach.  This significant disparity can be attributed to details of the multidimensional $\Phi$ ``landscape'' and to 
the innate differences between the two algorithms.  In CG optimization, minimization proceeds in a direction conjugate 
to the old gradient, $i.e.$, that the change in the function gradient be perpendicular to the most recent previous 
direction of minimization.  In contrast, the MINOP~\cite{De79} (``dogleg'') strategy is as follows: }

\renewcommand{\descriptionlabel}[1]%
         {\hspace{\labelsep}\textsf{#1}}
\begin{description}
\item[Step 0] Let $\Delta_i$ be a pre-set step bound, $\textbf{p}_i$ be the Cauchy step, and $\textbf{n}_i$ be the Newton 
step, $i.e.$, 
\begin{equation*}
 \begin{aligned}
  \textbf{p}_i &= (| \textbf{g}_i |^2 / \textbf{g}_{i}^{T} \textbf{G}_i \textbf{g}_i) \textbf{g}_i & \qquad \textbf{n}_i &= -\textbf{H}_i \textbf{g}_i
 \end{aligned}
\end{equation*}

\noindent
where $\textbf{g}_i$, $\textbf{G}_i$, and $\textbf{H}_i$ are the gradient, the Hessian approximation, and its inverse 
at the $i$th iteration respectively.

\item[Step 1] Compute $\textbf{n}_i$ and $| \textbf{n}_i |$. If $| \textbf{n}_i | \leq \Delta_i$, we take 
$\Delta \textbf{x}_i = \textbf{n}_i$.  
If $| \textbf{n}_i | > \Delta_i$, we compute $| t\textbf{n}_i |$, where $t$ is defined in Ref.~\cite{De79}. If 
$| t\textbf{n}_i | \leq \Delta_i$, we take $\textbf{x}_i = (\Delta_i / | \textbf{n}_i |)\textbf{n}_i$.

\item[Step 2] If $| t\textbf{n}_i | > \Delta_i$, we then compute the Cauchy step; and, if $| \textbf{p}_i | \geq \Delta_i$, we 
take $\Delta \textbf{x}_i = -(\Delta_i / | \textbf{g}_i |)\textbf{g}_i$.

\item[Step 3] If $| \textbf{p}_i | < \Delta_i$, we take $\Delta \textbf{x}_i = (1 - \theta_i)\textbf{p}_i + \theta_i t\textbf{n}_i$, 
where $\theta_i$ is chosen such that $0 < \theta_i < 1$ and $| \Delta \textbf{x}_i | = \Delta_i$.

\end{description}

\par{ The MINOP algorithm minimizes a real-valued function of any number of variables based on user-provided first 
derivative and function information.  In general, it applies a dogleg strategy which uses a gradient direction when 
one is far, a quasi-Newton direction when one is close, and a linear combination of the two when at intermediate 
distances from a solution.}

\newpage

\renewcommand{\thesection}{\arabic{section}}
\section{\label{sec:Section4}Results - Two Dimensions}
\hspace{17.5pt}

\numberwithin{equation}{section}

\par{ We have performed numerical simulations on a variety of system sizes, random initial configurations, numbers 
of constrained vectors ($i.e.$, $K$ cutoff), and choices for the independent parameters $D$ and $\alpha$.  Calculations 
have proceeded to attain high precision for the absolute minimum of the objective function of interest, $\Phi$ or 
$\tilde{\Phi}$.  Note that some initial configurations have not yielded the global minimum of the objective 
function hypersurface.  In such cases, it can be inferred that there exist some relative minima along the objective 
function's hypersurface landscape.  However, all cases that are reported below in our analysis involve the absolute 
minimum of the objective function.}

\par{ In order to simplify the presentation of our results, we introduce the following parameter

\begin{equation}
   \chi = \frac{M(K)}{dN}
\label{Equation17}
\end{equation}

\noindent
where $M(K)$ is defined as the number of independently constrained collective coordinates and $d$ indicates the system 
dimension.  The parameter $\chi$ is the ratio of the number of constrained degrees of freedom to the total number of 
degrees of freedom in the investigated system and has proved to be a fundamental descriptor in the prior one- and 
two-dimensional studies~\cite{Fa91,Uc04}.}

\par{ In the following two subsections, we present the results produced by application of the MINOP algorithm to 
minimization of the objective function $\Phi$ for two-dimensional particle systems.  In the first subsection, 
we demonstrate the ability to tailor the small-$|\boldsymbol{k}|$ portion of the structure factor associated with 
point particle systems.  In the latter subsection, we study the effect of manipulating the structure factor $S(k)$ 
within the distinctive ``wavy crystalline'' regime reported previously in Ref.~\cite{Uc04}.}


\renewcommand{\thesection}{\arabic{section}}
\subsection{\label{sec:Section4a}Tailoring the Structure Factor - Two Dimensions}
\hspace{17.5pt}

\par{ This section describes our successful attempts to manipulate the structure factor $S(k)$ of  
two-dimensional point particle systems.  We subjected the objective function $\Phi$ with a randomly 
generated initial configuration to the MINOP algorithm in order to evolve the $C(\boldsymbol{k})$'s 
toward their target values.  Our numerical simulations have involved the use of 168- and 418-particle systems.  
Note that these system sizes are the same as those studied in Ref.~\cite{Uc04}.  In addition, we vary 
the $\boldsymbol{k}$-space range parameter $K$ ($2\pi$ $\leq K \leq$ $40\pi$) and the multiplicative parameter 
$D$ (0.01 $\leq D K^{\alpha} \leq$ 150).  We have varied $\alpha$, a key parameter determining the nature of 
the tailored structure factor.  In particular, we have generated solutions for the Harrison-Zeldovich~\cite{Ga02} 
spectrum ($\alpha$ = 1) as well as for the $\alpha$ = 4, 6, 8, and 10 minimization problems.  We devoted much of 
our study to the $\alpha$ = 1 and 6 minimization problems with less detailed attention to the 
$\alpha$ = 4, 8, and 10 minimization cases.  As discussed below, our results demonstrate wide latitude in the capacity 
to control the structure factor for two-dimensional point particle systems.}

\begin{figure}[H]
\begin{center}
\includegraphics[width=.4\linewidth, angle=270, clip=]{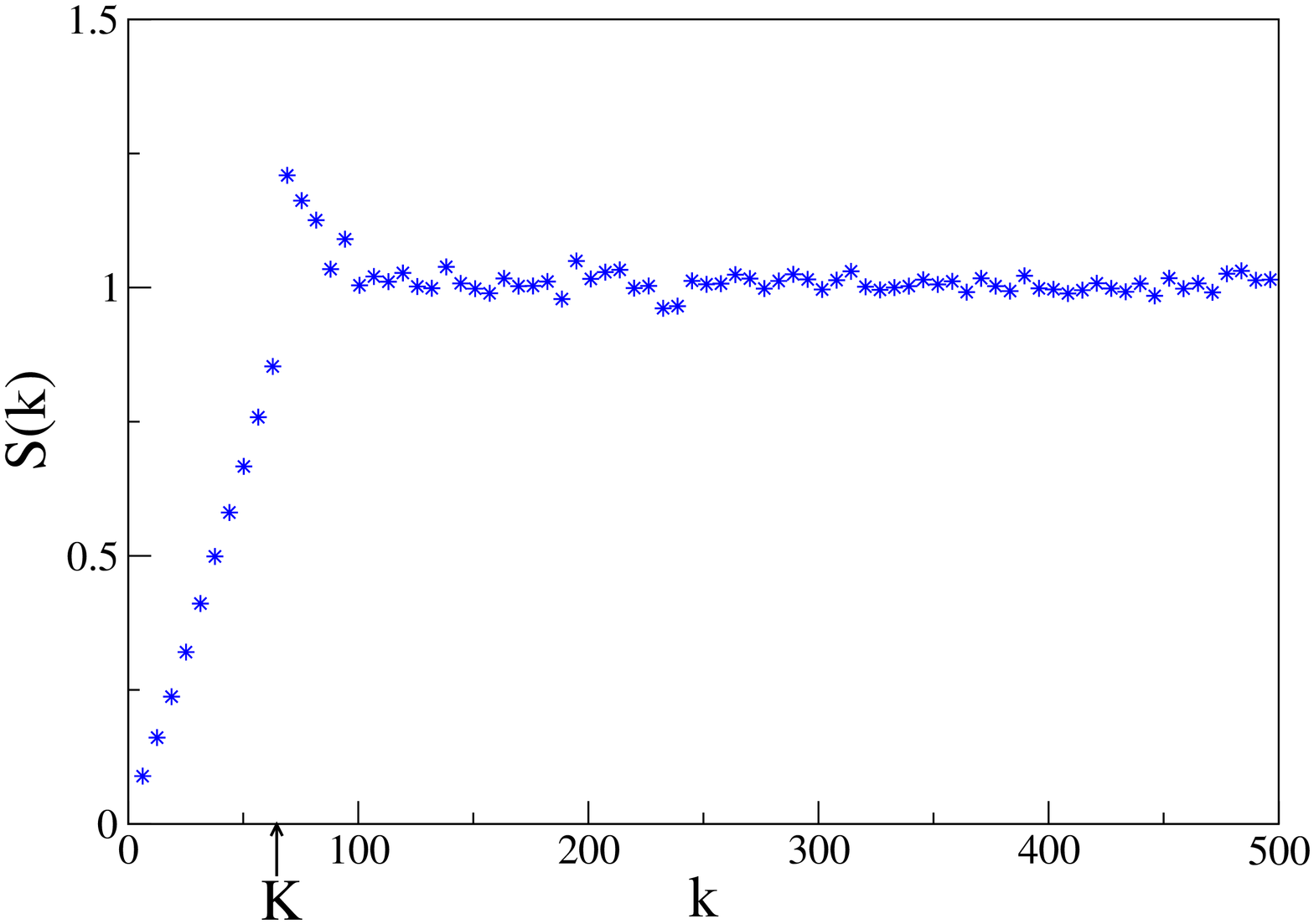}
\vspace{0.2in}
\includegraphics[width=.4\linewidth, angle=270, clip=]{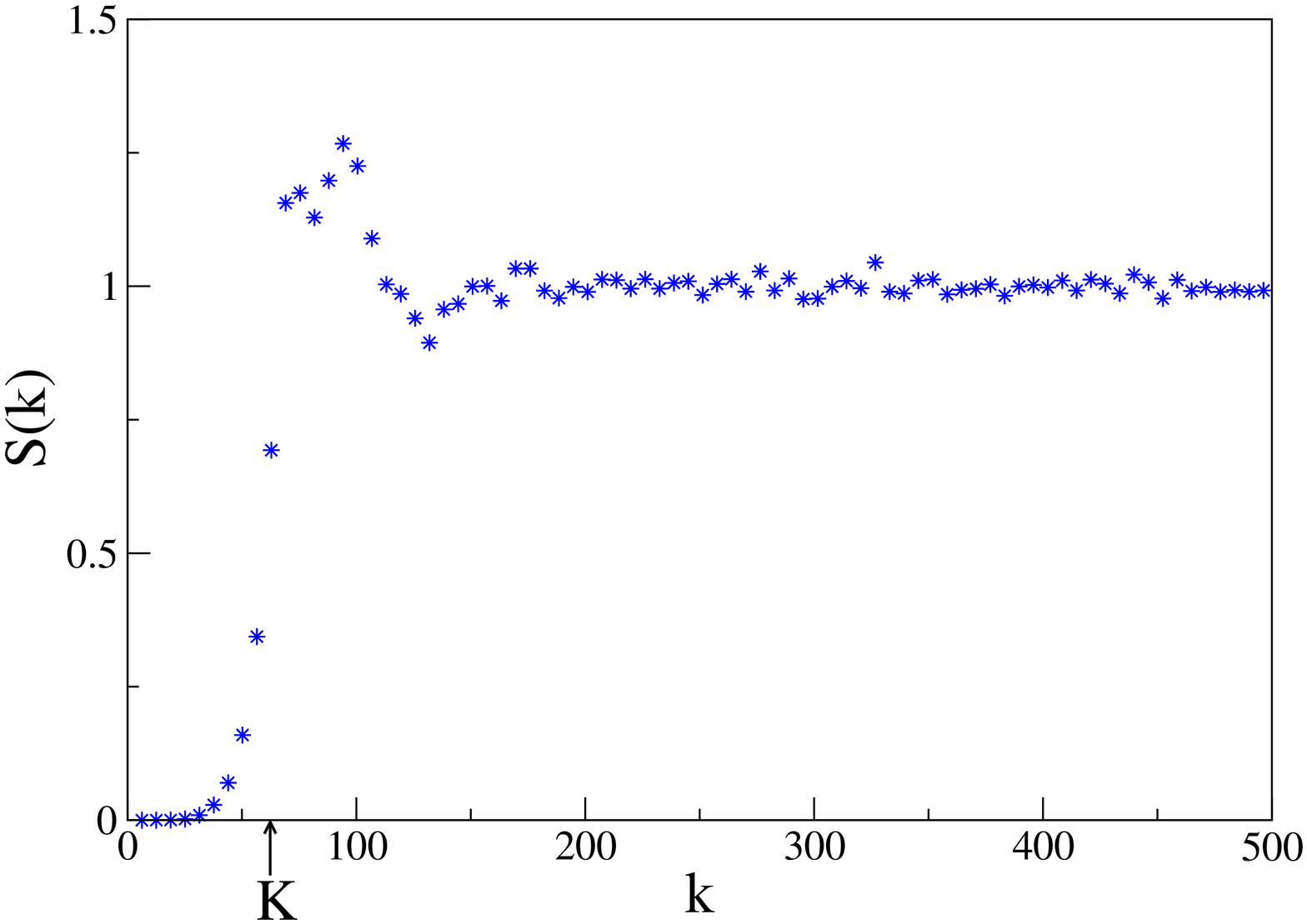}
\end{center}
\caption{ (Color online) Averaged structure factor plots for the two-dimensional minimization problem.  The relevant 
parameters are $N$ = 168, $K$ = 20$\pi$, $D K^{\alpha}$ = 75, $\chi$ = 0.470238.  Top panel: 
Harrison-Zeldovich linear spectrum for small k.  Bottom panel: $|\boldsymbol{k}|^6$ spectrum.  Each 
structure factor is averaged over 50 realizations.} 
\label{fig:Sk2Dk1+Sk2Dk6}
\end{figure}

\par{ Plots of the system structure factor that have been tailored to fit the linear Harrison-Zeldovich and $|\boldsymbol{k}|^6$ 
spectra are shown in Fig.~\ref{fig:Sk2Dk1+Sk2Dk6}.  Here, the structure factor is derived from final configurations 
via Eq.~\ref{Equation2} and averaged over 50 independent realizations.  For the purposes of graphical representation, 
the structure factor is binned over the reported range of $\boldsymbol{k}$-space.  Note that by construction, the 
contributions to $S(k)$ below the cutoff $K$ for a specific set of parameters (fixed $N$, $K$, $D$, and $\alpha$) 
are identical for each realization.  Outside the cutoff, the $S(k)$ contributions from each realization deviate 
irregularly from one another, but their average for $|\boldsymbol{k}| > K$ shows a weak maximum followed by quick 
decay to unity as $|\boldsymbol{k}|$ increases.  From these structure factor plots, it is evidently possible to 
tailor the structure factor to either spectrum.  In particular, the respective linear and sextic nature of the plots 
at low $k$ are visually clear.}

\begin{figure}[H]
\begin{center}
\fbox{\includegraphics[width=.3\linewidth]{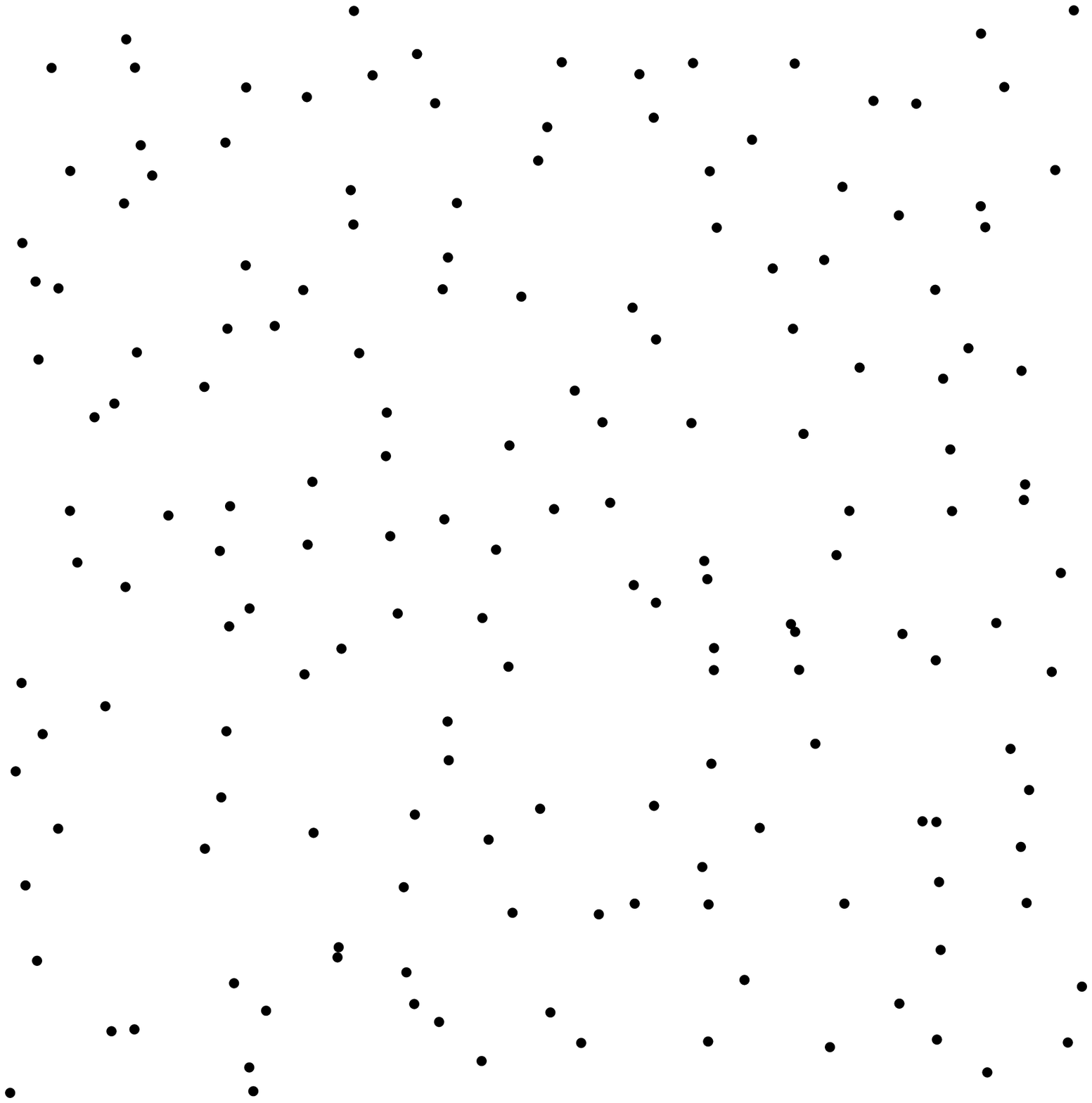}}
\end{center}
\vspace{0.1in}
\begin{center}
\fbox{\includegraphics[width=.3\linewidth]{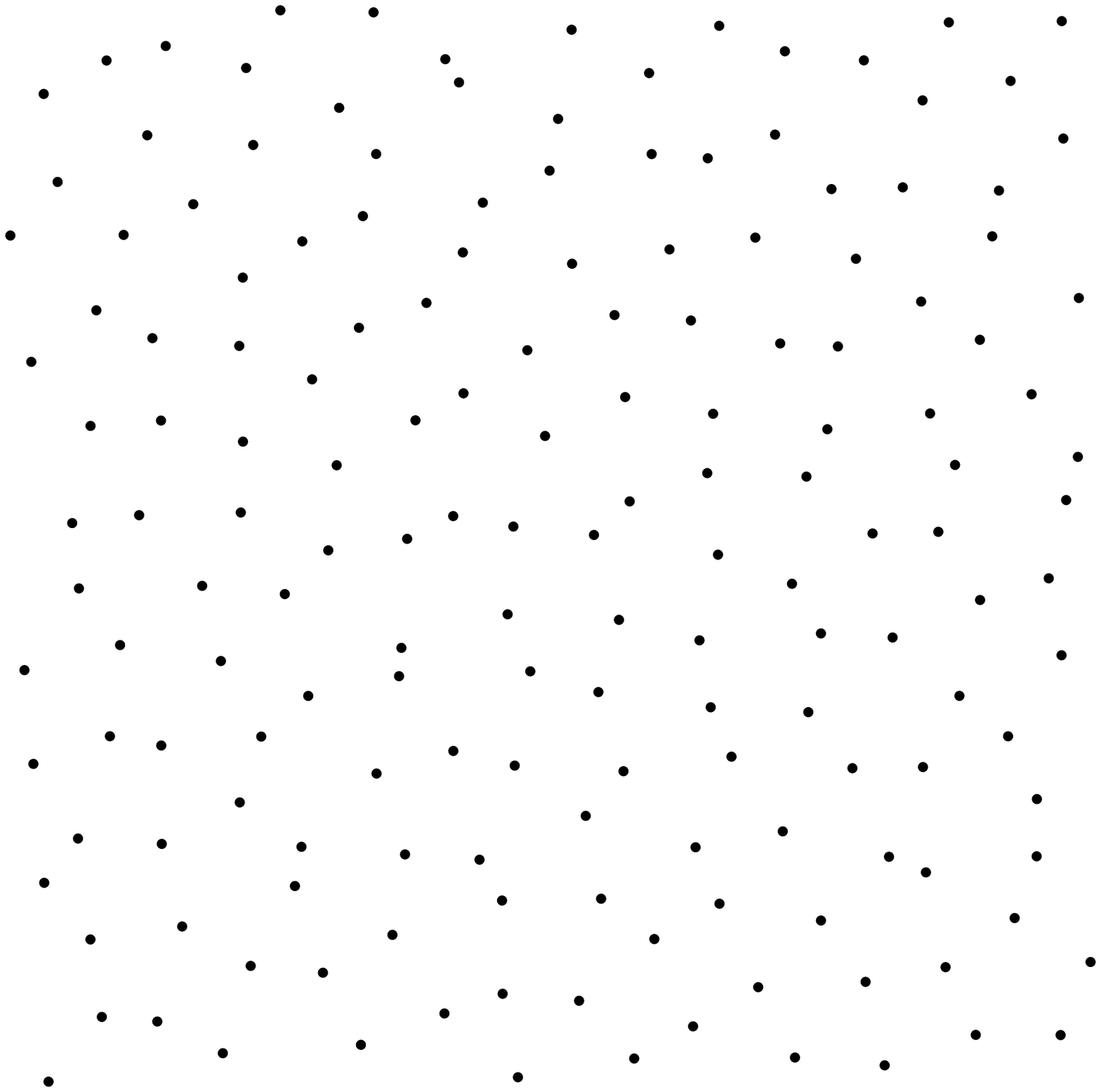}}
\end{center}
\caption{ Typical two-dimensional final configurations for the investigated spectra.  The relevant parameters 
are $N$ = 168, $K$ = 20$\pi$, $D K^{\alpha}$ = 75, $\chi$ = 0.470238.  Top panel: Harrison-Zeldovich 
linear spectrum.  Bottom panel: $|\boldsymbol{k}|^6$ spectrum.}
\label{fig:C2Dk1+C2Dk6}
\end{figure}

\par{ Figure~\ref{fig:C2Dk1+C2Dk6} displays sample two-dimensional final configurations generated by the $\Phi$ minimization 
problem for the linear Harrison-Zeldovich and $|\boldsymbol{k}|^6$ spectra.  The lack of any visible long-range 
regularity is apparent in the pictured configurations for both cases.  However, some random point clustering appears to be 
present in the accompanying configuration for the Harrison-Zeldovich spectrum.  This differs from the $|\boldsymbol{k}|^6$ 
spectrum configuration in which an effective repelling particle core appears to be present.  Finally, we mention that 
corresponding analysis of the $|\boldsymbol{k}|^4$, $|\boldsymbol{k}|^8$, and $|\boldsymbol{k}|^{10}$ spectra 
minimization problems yield results consistent with the above observations, providing systematic pattern sequences in 
both $\boldsymbol{r}$ and $\boldsymbol{k}$ spaces as illustrated in Figs.~\ref{fig:Sk2Dk1+Sk2Dk6} and~\ref{fig:C2Dk1+C2Dk6}.}

\renewcommand{\thesection}{\arabic{section}}
\subsection{\label{sec:Section4b}Effect of $|\boldsymbol{k}|^{6}$ Spectrum Imposition on the Wavy Crystalline Regime}
\hspace{17.5pt} 

\par
\begin{figure}[H]
\begin{center}
\fbox{\includegraphics[width=.3\linewidth]{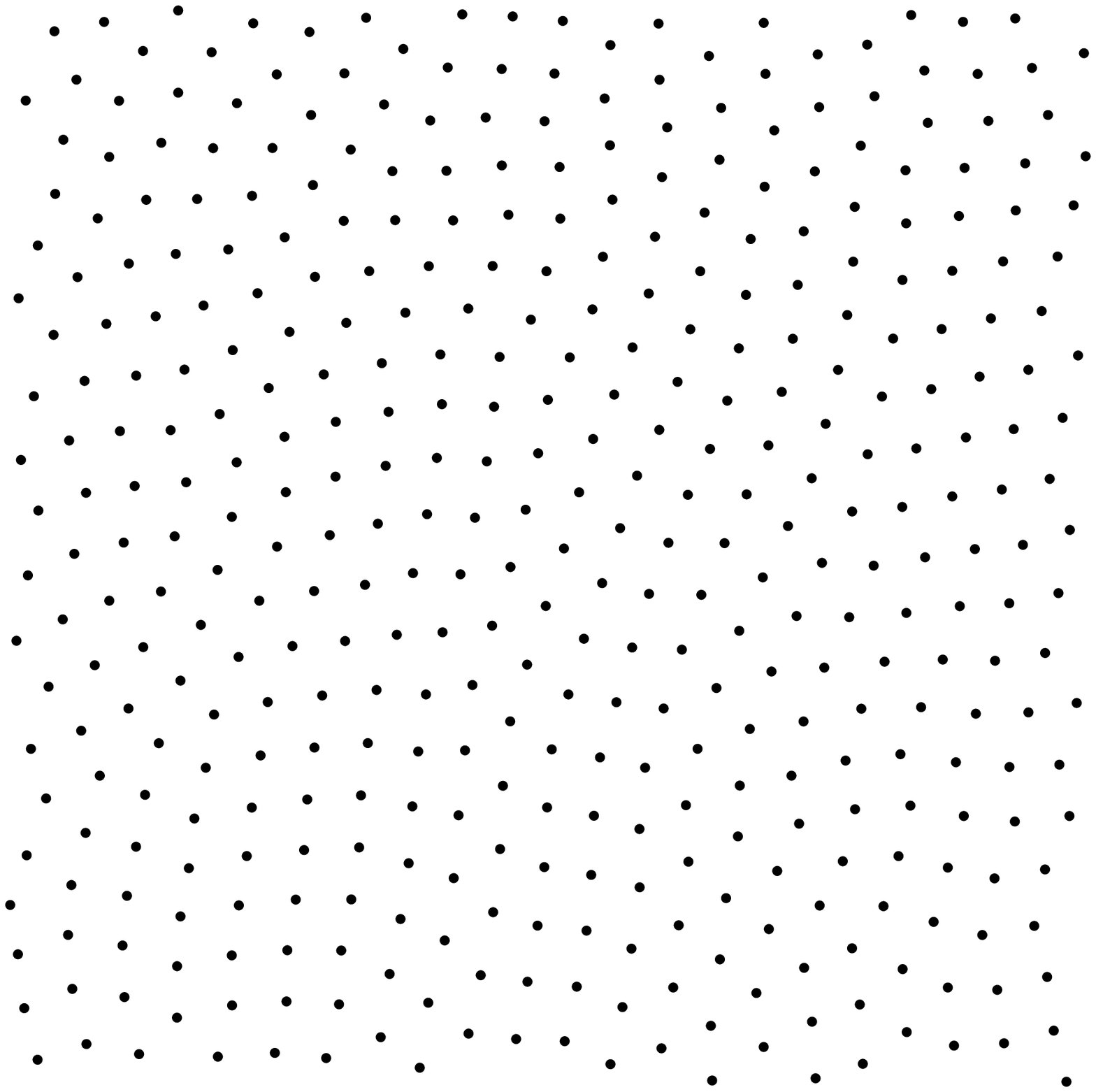}}
\end{center}
\vspace{0.1in}
\begin{center}
\fbox{\includegraphics[width=.3\linewidth]{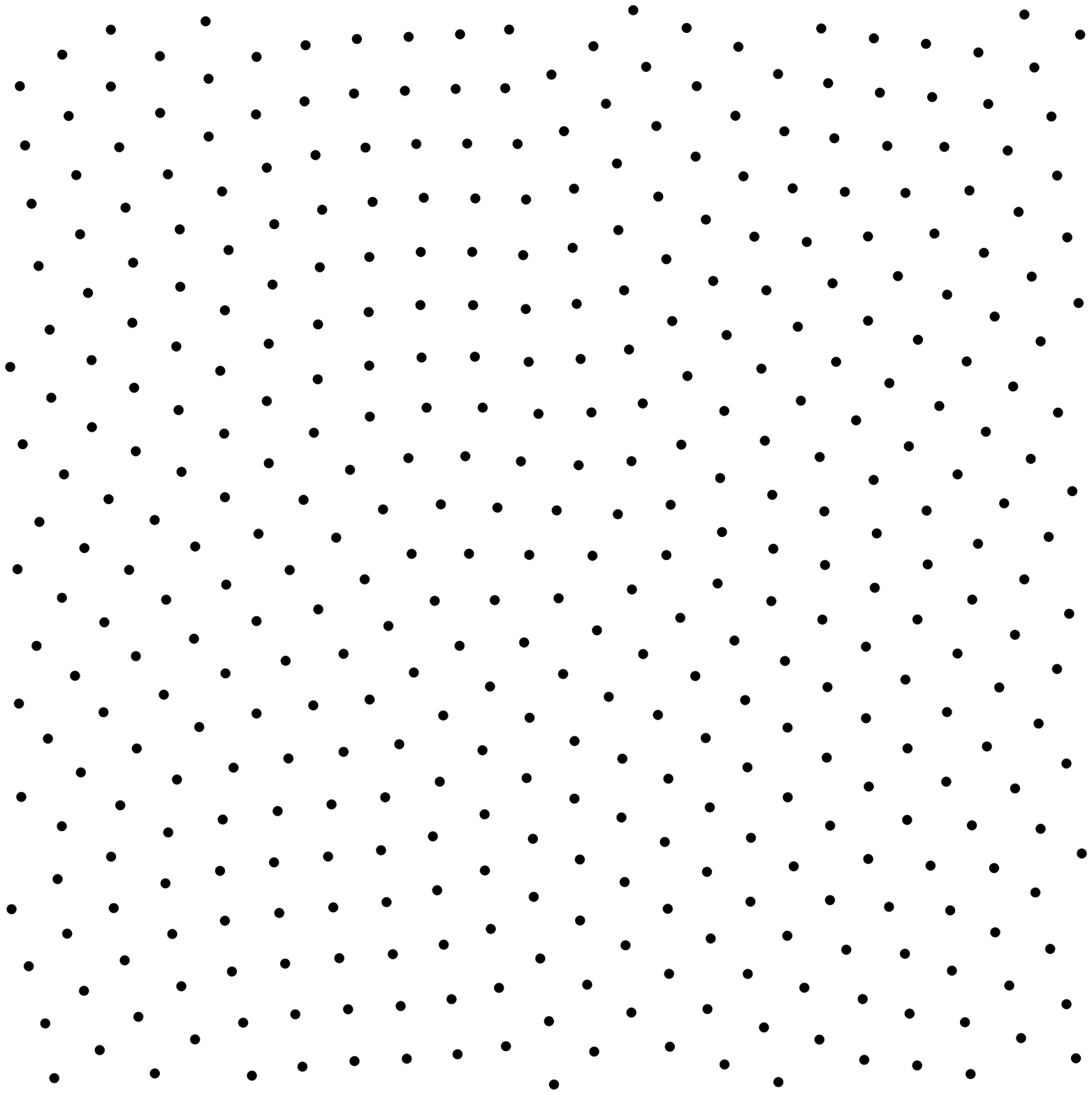}}
\end{center}
\vspace{0.1in}
\begin{center}
\fbox{\includegraphics[width=.3\linewidth]{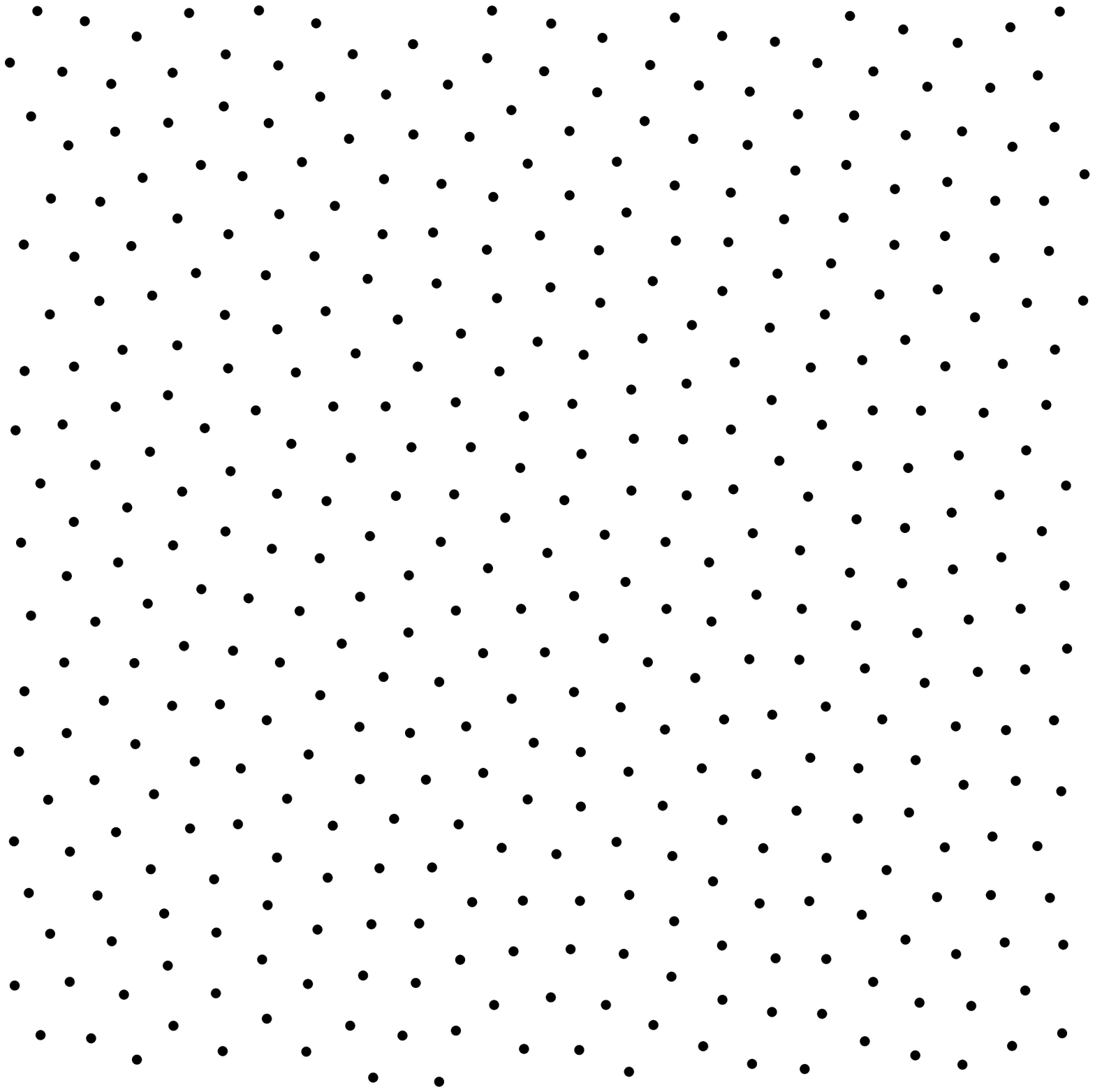}}
\end{center} 
\caption{ Real space particle patterns for a system of 418 point particles.  The $C(\boldsymbol{k})$ quantities for the 
wave vectors consistent with $\chi$ = 0.581339 have been constrained.  Top panel: Particle pattern in the wavy crystalline 
regime ($D$ = 0).  Middle panel: Particle pattern on which the $|\boldsymbol{k}|^6$ spectrum has been imposed with 
$D K^6$ = 0.01.  Bottom panel: Particle pattern on which the $|\boldsymbol{k}|^6$ spectrum has been imposed with $D K^6$ = 10.}
\label{fig:WCk0+WCk61+WCk62}
\end{figure}

{ One of the qualitatively distinct regimes that was isolated in our earlier two-dimensional collective coordinates 
study~\cite{Uc04} was the ``wavy crystalline'' regime.  This regime is distinguished by patterns consisting of 
particle columns that display a meandering displacement away from linearity.  For $N =$ 418, we have found the patterns 
to occur when 0.58 $\leq \chi \leq$ 0.78~\cite{Uc04}.  An example of such a pattern is displayed in the top panel of 
Fig.~\ref{fig:WCk0+WCk61+WCk62}.  The result of applying the $|\boldsymbol{k}|^6$ spectrum minimization to this 
$\chi$ interval is our focus in this section.  As above, we subjected randomly generated particle systems 
to the MINOP~\cite{De79} algorithm to find the absolute minimum of the objective function $\Phi$.  Our $\boldsymbol{k}$-space 
parameter $K$ is determined by the $\chi$ range over which the wavy crystalline regime prevails.  While focusing on the 
$|\boldsymbol{k}|^6$ spectrum minimization problem, we have varied the independent coefficient $D$.}

\par{ The middle and bottom panels of Fig.~\ref{fig:WCk0+WCk61+WCk62} display 418-particle configurations that result from 
imposing the $|\boldsymbol{k}|^6$ spectrum on particle systems that at $D=$ 0 lie in the wavy crystalline regime.  The 
multiplicative parameter $D$ used in generating the configurations in the middle and bottom panels is low ($D K^6=$ 0.01) 
and high ($D K^6=$ 10), respectively, and the cutoff $K$ is near the lower limit of the wavy regime of interest.  It is 
important to note that all three particle patterns in Fig.~\ref{fig:WCk0+WCk61+WCk62} were formed from a common initial 
configuration.  A comparison of the three figures reveals distortion and disruption of the meandering nature of the 
reference ($D=$ 0) configuration that intensifies as $D$ increases.  Evidently the previously documented~\cite{Uc04} 
tendency at $D =$ 0 for increasing $\chi$ to herd particles towards a crystalline arrangement is sabotaged by allowing 
$D$ to increase.}

\renewcommand{\thesection}{\arabic{section}}
\section{\label{sec:Section5}Results - Three Dimensions}
\hspace{17.5pt}

\numberwithin{equation}{section}

\par{ In this section, we discuss results for three-dimensional configurations that have been subjected to the 
computational algorithm, MINOP~\cite{De79} requiring absolute minimization of $\Phi$ and $\tilde{\Phi}$.  First, we 
extend our two-dimensional analysis to the tailoring of the structure factor of three-dimensional particle systems.  
Second, we return to an examination of the configurational patterns associated with constrained collective density 
variables in three dimensions.  This latter aspect extends our earlier work in one~\cite{Fa91} and two~\cite{Uc04} 
dimensions.  Note that all calculations presented in this section have been carried out to the same high precision 
as for the two-dimensional samples discussed in the preceding Sec.~\ref{sec:Section4}.}

\renewcommand{\thesection}{\arabic{section}}
\subsection{\label{sec:Section5a}Tailoring the Structure Factor $S(k)$ - Three Dimensions}
\hspace{17.5pt}

\par{ Our numerical calculations minimizing $\Phi$ have involved 256- and 500-particle systems.  Note that these 
system sizes conform to the guideline discussed in Sec.~\ref{sec:Section3}.  We vary the $\boldsymbol{k}$-space range 
parameter $K$ ($2\pi$ $\leq K \leq$ $40\pi$), the multiplicative parameter $D$ (0.01 $\leq D K^{\alpha} \leq$ 180), and 
investigate both the $\alpha =$ 1 Harrison-Zeldovich spectrum and the $\alpha =$ 6 minimization problem.}

\begin{figure}[H]
\begin{center}
\includegraphics[width=.4\linewidth, angle=270, clip=]{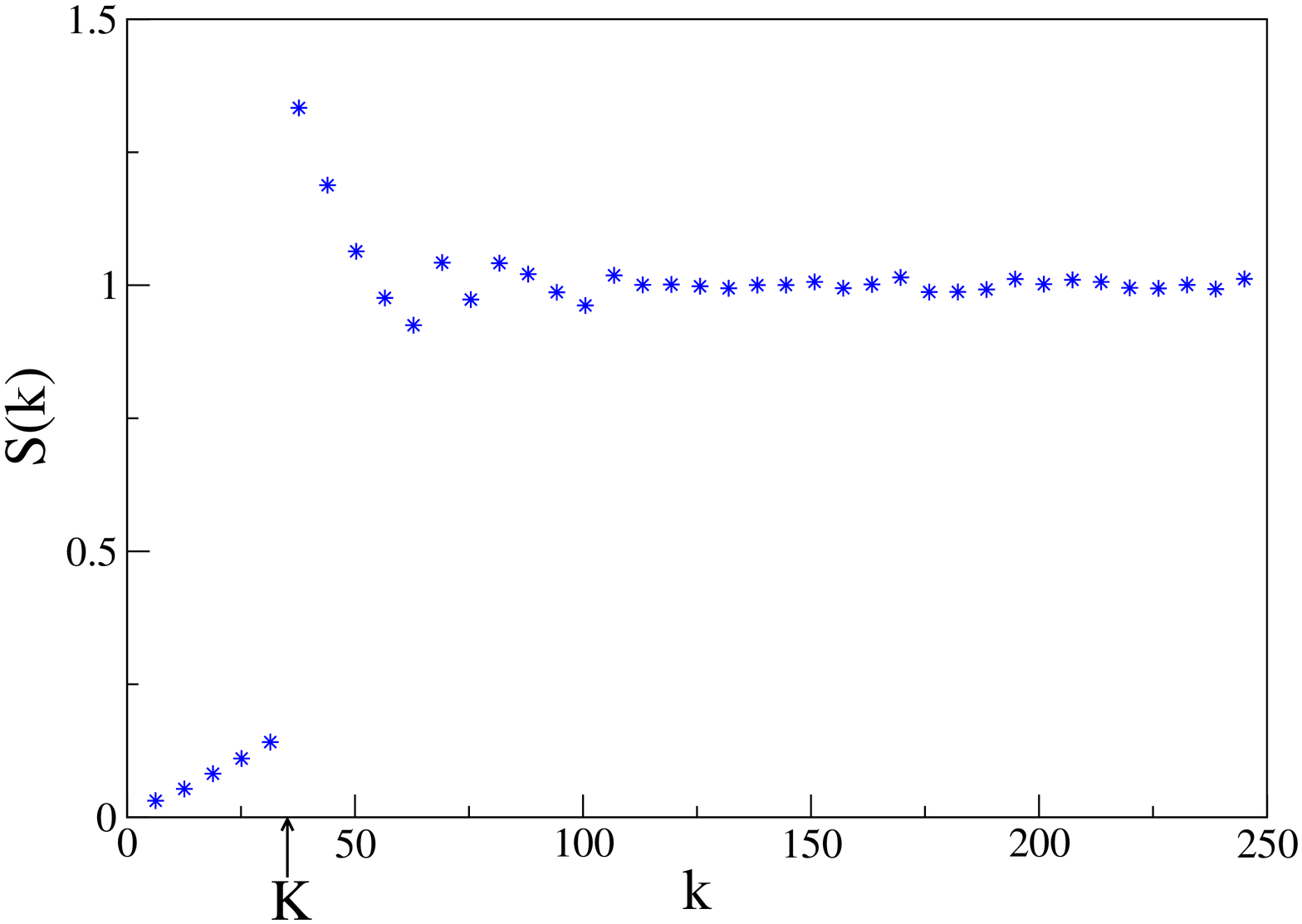}
\vspace{0.2in}
\includegraphics[width=.4\linewidth, angle=270, clip=]{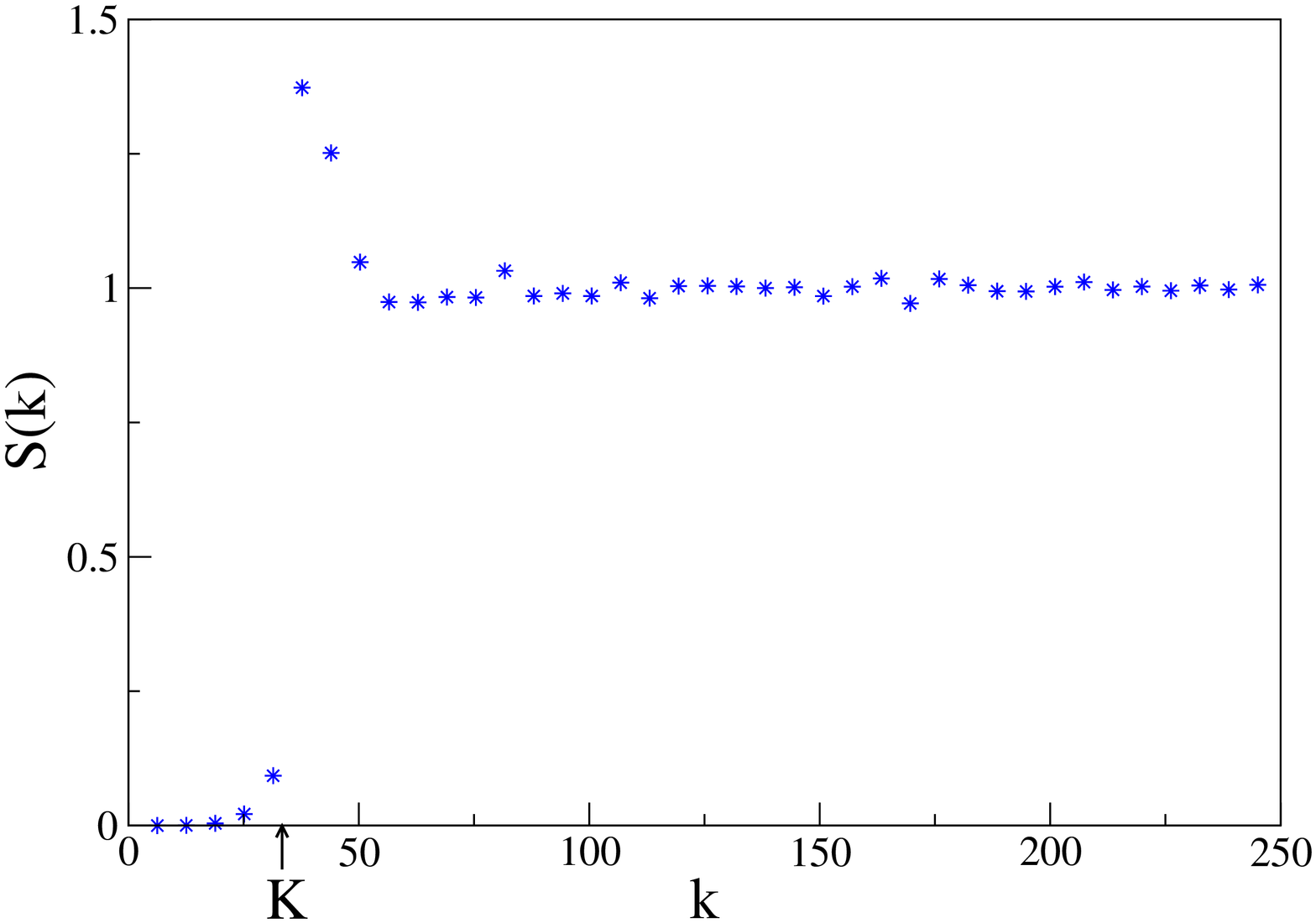}
\end{center}
\caption{ (Color online) Averaged structure factor plots for the three-dimensional minimization problem.  The relevant 
parameters are $N$ = 256, $K$ = 10$\pi$, $D K^{\alpha}$ = 20, $\chi$ = 0.334635.  Top panel: 
Harrison-Zeldovich spectrum.  Bottom panel: $|\boldsymbol{k}|^6$ spectrum.  Each structure factor is 
averaged over 6 realizations.}
\label{fig:Sk3Dk1+Sk3Dk6}
\end{figure}

\begin{figure}[H]
\begin{center}
\fbox{\includegraphics[width=.4\linewidth]{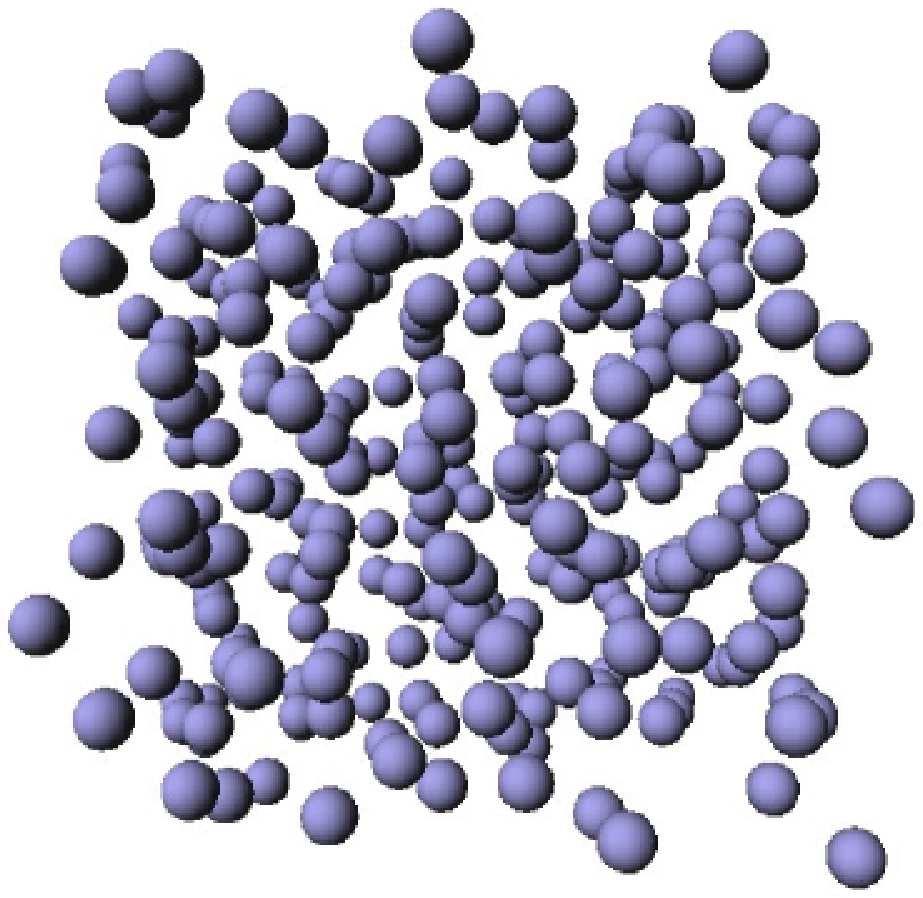}}
\end{center}
\vspace{0.1in}
\begin{center}
\fbox{\includegraphics[width=.4\linewidth]{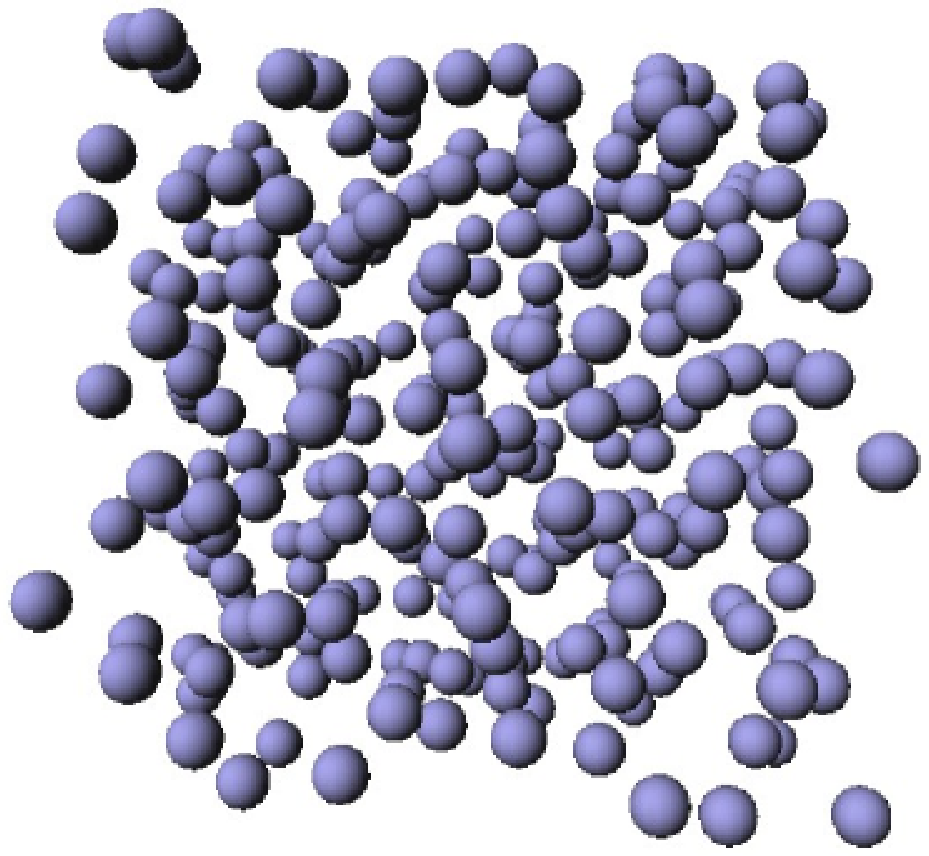}}
\end{center}
\caption{ (Color online) Snapshots of three-dimensional final configurations for investigated spectra.  The relevant 
parameters are $N$ = 256, $K$ = 10$\pi$, $D K^{\alpha}$ = 20, $\chi$ = 0.334635.  Top panel: 
Harrison-Zeldovich spectrum.  Bottom panel: $|\boldsymbol{k}|^6$ spectrum.}
\label{fig:C3Dk1+C3Dk6}
\end{figure}

\par{ As before, structure factors are derived from final configurations averaged over several realizations.  The 
linear Harrison-Zeldovich and $|\boldsymbol{k}|^6$ cases are displayed in Fig.~\ref{fig:Sk3Dk1+Sk3Dk6}.  Similarly 
to the two-dimensional study, the tailored structure factor deviates irregularly for $|\boldsymbol{k}|$ $>$ $K$ 
prior to averaging over a set of realizations.  Once again the averaged $S(k)$ exhibits a peak just beyond 
$|\boldsymbol{k}| = K$, now an even stronger feature than in the two-dimensional cases shown earlier in 
Fig.~\ref{fig:Sk2Dk1+Sk2Dk6}.  An examination of the plotted structure factors clearly reveals the linear and 
sextic nature of the two curves near the origin.  Figure~\ref{fig:C3Dk1+C3Dk6} displays representative 
three-dimensional final configurations for the $\Phi$ minimization problem for both the linear Harrison-Zeldovich 
and $|\boldsymbol{k}|^6$ spectra.  The high degree of disorder is evident in both sample configurations.}

\par{ A system-size scaling study was performed as part of our investigation.  This is relevant to the approach 
to the thermodynamic limit.  Specifically, we compared the simulation times for two 
different system sizes of fixed $\chi$ (appropriately scaled independent parameters $N$, $K$, and $D$) for the 
Harrison-Zeldovich spectrum.  We find that doubling the system size increases the computation time by approximately a 
factor of 10.}

\renewcommand{\thesection}{\arabic{section}}
\subsection{\label{sec:Section5b}Minimizing Collective Density Variables $C(\boldsymbol{k})$ - Three Dimensions}
\hspace{17.5pt}

\par{ Our numerical studies for $D =$ 0 concentrated on two system sizes ($N$ = 108 and 500), a wide range of 
$\boldsymbol{k}$-space constraints $i.e.$ low through high $K$, and a variety of initial configurations (both random 
and lattice-based).  For the wide range of $\boldsymbol{k}$-space traversed, the odds of hitting relative minima 
of the total energy function $\tilde{\Phi}$ was increased for large $K$.  However, trajectories converged to the 
absolute $\tilde{\Phi}$ minimum for all cases that were included in our final analysis.  Our findings remain 
substantially consistent between the two investigated system sizes.}

\hspace{0.1in}

\begin{table}[ht]
\caption{\label{tab:Table1} Classification of investigated cases associated with each of the regimes.  Note that 
the multiplicative parameter $D =$ 0 for this objective function $\tilde{\Phi}$ minimization problem.}
\begin{ruledtabular}
\begin{tabular} {ccc}   
$N$  &  \multicolumn{1}{c}{Disordered Regime}   &   \multicolumn{1}{c}{Crystalline Regime}  \\ \hline  
108  &       $\chi \leq$ 0.469136   &        $\chi \geq$  0.524691     \\
500  &       $\chi \leq$ 0.500666   &        $\chi \geq$  0.516666     \\
\end{tabular}
\end{ruledtabular}
\end{table}

\par{ In the course of our investigation, we observed two distinct regimes of the final configurations as $\chi$ 
(see Eq.~\ref{Equation17}) was varied: disordered and crystalline.  Table~\ref{tab:Table1} presents the relevant 
range in $\chi$ for the two investigated systems and distinguishable regimes.  The table indicates disorder 
for low values of $\chi$ and crystallinity for high values of $\chi$.  This is expected by analogy with the 
reported results of the one-~\cite{Fa91} and two-dimensional~\cite{Uc04} articles on collective density 
variables.  However, our three-dimensional analysis indicates an abrupt transition from disordered to crystalline 
regimes revealing the lack of an intermediate ``wavy crystalline'' regime as observed in two dimensions.
For the 500-particle system and $\chi \geq$  0.516666, we have verified that the crystal structure
is a face-centered cubic lattice, which for the density of the system is consistent with the predictions of S\"{u}t\'{o}
\cite{Su05}.}

\begin{figure}[H]
 \begin{center}
 \fbox{\epsfig{file=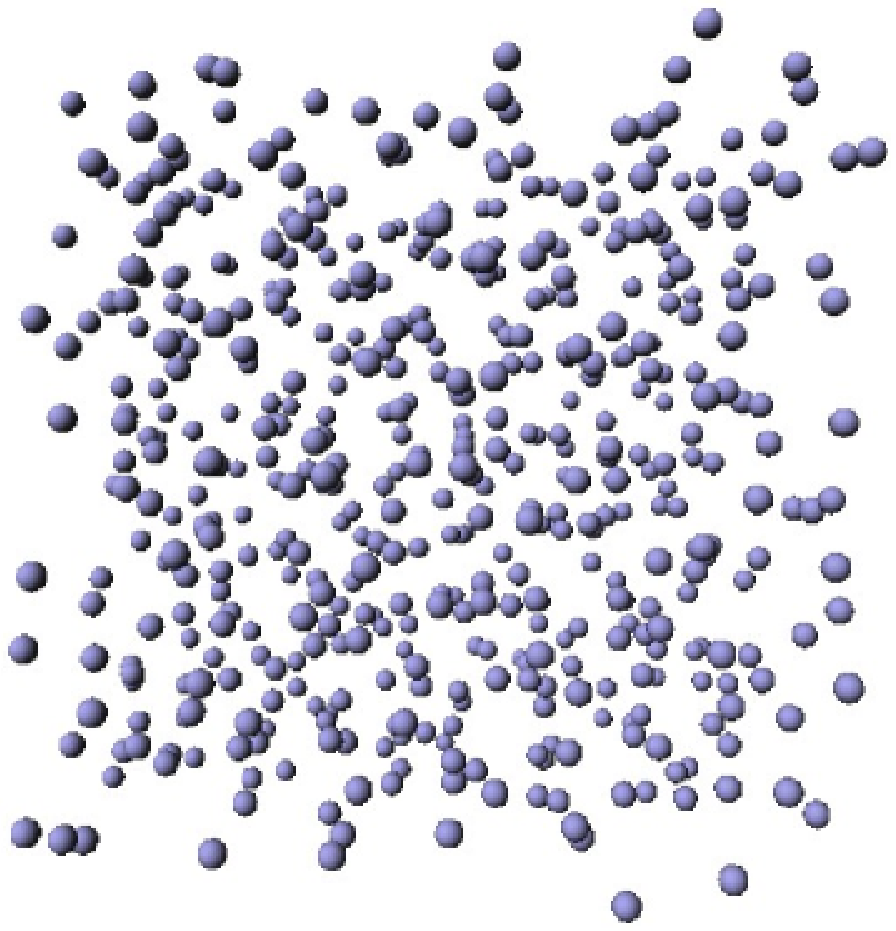, width=.45\linewidth, clip=}}
 \end{center}
 \vspace{0.1in}
 \begin{center}
 \fbox{\epsfig{file=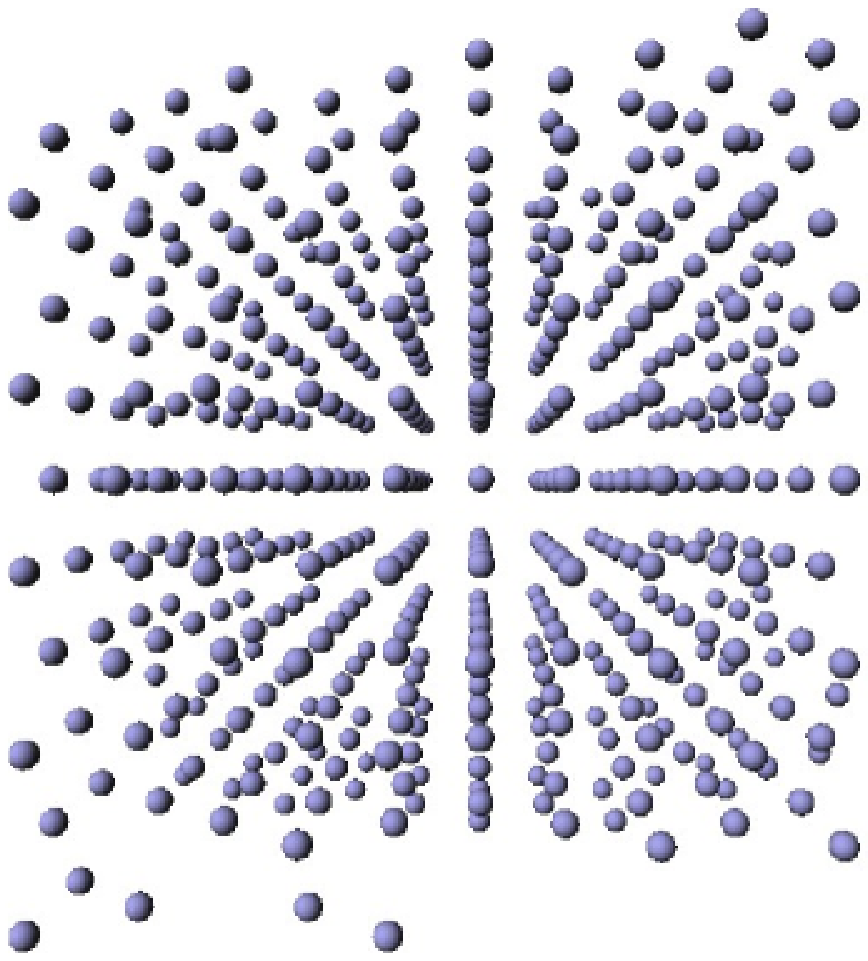, width=.45\linewidth, clip=}}
 \end{center}
\caption{ (Color online) Real space particle patterns for the two distinct regimes for systems of 500 point particles.  Top panel: 
Particle pattern in the disordered regime.  The $C(\boldsymbol{k})$ quantities consistent with parameter $\chi$ = 0.171333 
have been constrained to their minimum values $-N$/2.  Bottom panel: Particle pattern in the crystalline regime.  The 
$C(\boldsymbol{k})$ quantities consistent with parameter $\chi$ = 0.702666 have been constrained to their minimum values 
$-N$/2.}
\label{fig:C3D1+C3D2}
\end{figure}

\begin{figure}[H]
\begin{center}
\includegraphics[width=.4\linewidth, angle=270, clip=]{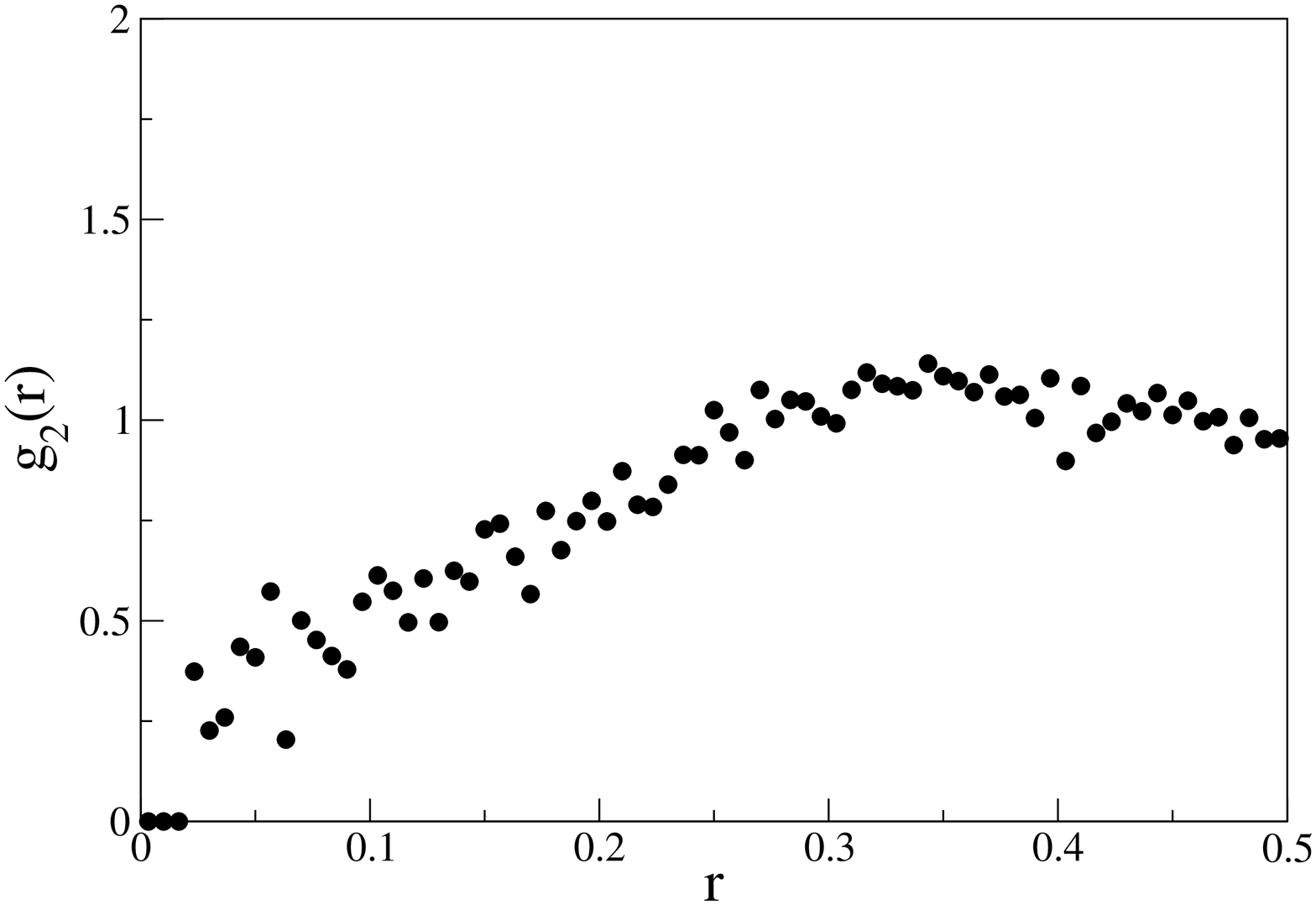}
\vspace{0.2in}
\includegraphics[width=.4\linewidth, angle=270, clip=]{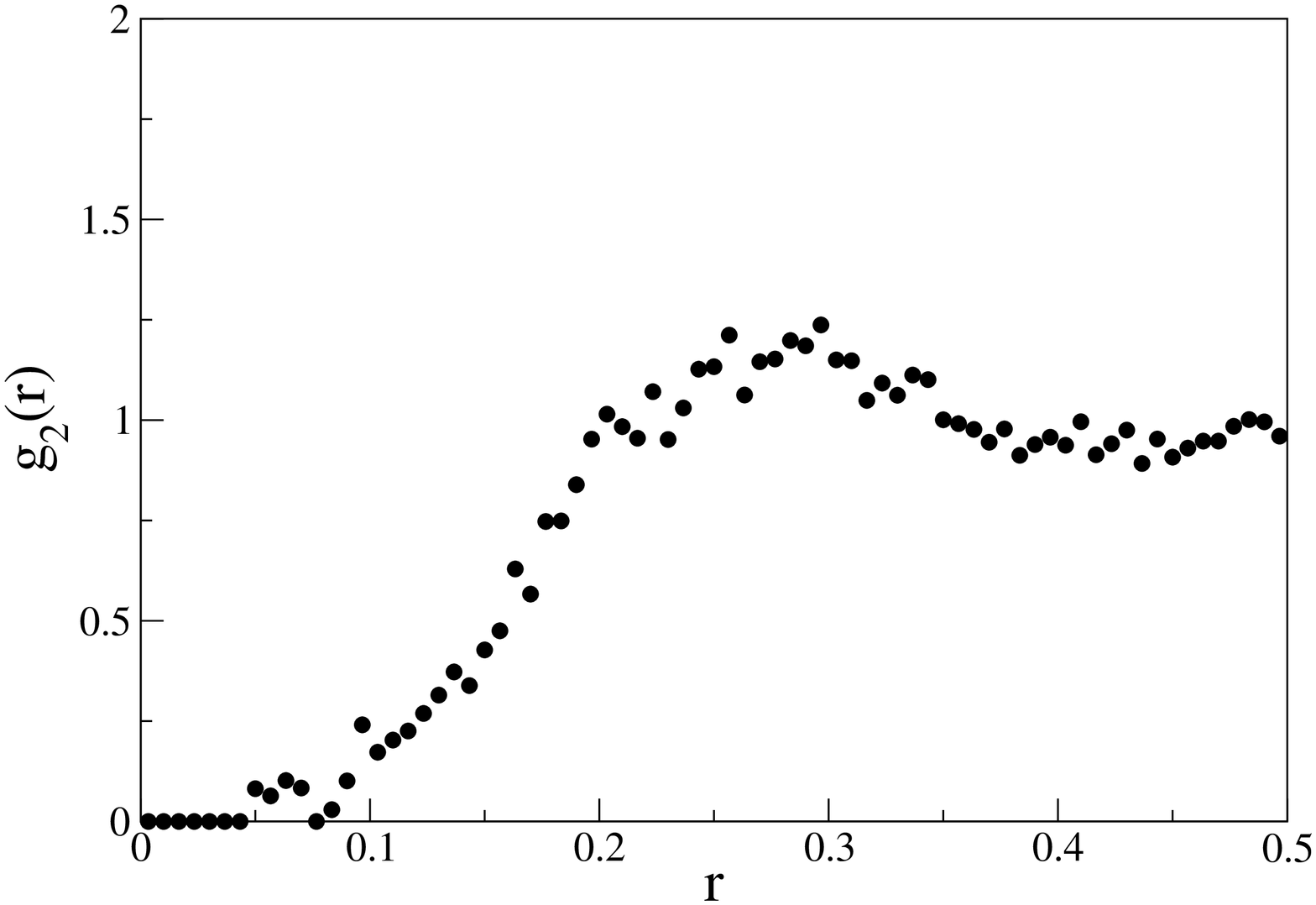}
\end{center}
\caption{ Radial distribution functions for systems of 108 point particles. Top panel: The $C(\boldsymbol{k})$ quantities 
consistent with parameter $\chi$ = 0.123457 have been constrained to their minimum values $-N$/2.  Bottom panel: The 
$C(\boldsymbol{k})$ quantities consistent with parameter $\chi$ = 0.262346 have been constrained to their minimum values 
$-N$/2.  Each radial distribution function is averaged over 10 realizations.}
\label{fig:G23D1+G23D2}
\end{figure}

\par{ Particle patterns respectively within the disordered (top panel) and crystalline (bottom panel) regimes for 
$N$ = 500 are displayed in Fig.~\ref{fig:C3D1+C3D2}.  The examples shown are typical for the two regimes and are 
vividly distinct.  Further insight into the generated point patterns follows from examination of the associated pair 
correlation functions~\cite{TB02} (see Fig.~\ref{fig:G23D1+G23D2}).  The emergence of an effective repulsive core 
for increasing values of $\chi$ is apparent, reminiscent of a similar effect observed in Ref.~\cite{Uc04} in the 
two-dimensional case.}


\renewcommand{\thesection}{\arabic{section}}
\section{\label{sec:Section6}Conclusions and Discussion}
\hspace{17.5pt}

\numberwithin{equation}{section}

\par{ The former studies of collective coordinate properties presented in Refs.~\cite{Fa91} and~\cite{Uc04} were 
restricted to one- and two-dimensional point patterns, and documented the effect of forcing sets of the collective 
variables for $|\boldsymbol{k}| \leq K$ to their individual absolute minima.  The present investigation extends 
that analysis in two distinct directions, by considering point patterns in three dimensions, and by examining the 
effect of constraining the collective variables around the $\boldsymbol{k}$-space origin to chosen increments above 
their absolute minima.  Specific assignments of the increments that have been considered have the form 
$D|\boldsymbol{k}|^{\alpha}$, where $D >$ 0, and $\alpha =$ 1, 4, 6, 8, and 10.  Point particle configurations 
satisfying these various collective variable constraints have been obtained to high numerical precision starting from 
both random and from distorted-crystal initial conditions, followed by minimization of appropriate objective functions 
[$\Phi$ and $\tilde{\Phi}$, Eqs.~\ref{Equation7} and~\ref{Equation13}].  Although previous attempts have produced 
configurations for which the associated structure factor $S(k) \propto k^{\alpha}$ with $\alpha \leq 4$~\cite{Ga02,Ga04}, 
our analysis also yields specific configurations that exhibit $S(k) \propto k^{\alpha}$ with $\alpha \geq 6$.  In addition 
to the specific cases reported in this paper, our studies reveal that a considerably wider range of constrained 
$C(\boldsymbol{k})$ patterns can be imposed on many-particle systems.  For example, forcing $C(\boldsymbol{k})$ to 
equal 0 instead of its absolute minimum $-N$/2, reveals a tendency to produce a high degree of disorder in the 
generated particle systems.}

\par{ When $D =$ 0, increasing the fraction $\chi$ of system degrees of freedom subject to collective variable constraints 
was found in one and two dimensions to drive the point particle configurations more and more toward crystalline periodicity.  
The present extension not surprisingly shows that the same qualitative trend applies to three dimensions as well.  However 
the ``wavy crystalline'' regime reported in Ref.~\cite{Uc04} to separate the two-dimensional low-$\chi$ disordered regime 
from the high-$\chi$ crystalline regime appears to have no three-dimensional analog.  As $\chi$ increases for $D =$ 0 cases 
in three dimensions, the relevant $\tilde{\Phi}$ landscape on which numerical absolute minimization needs to be carried out 
develops an increasing density of relative minima, thereby inhibiting (but not necessarily preventing) the search for qualifying 
particle configurations.  For the few cases considered, increasing $D$ above zero at constant $\chi$ has the effect of 
disrupting the tendency toward crystalline order.}

\par{ A characteristic feature of the various cases examined, provided that the fraction $\chi$ is below its maximum value 
to yield valid solutions, is degeneracy of the final particle configurations.  This is obvious when several independent 
initial configurations for a case considered are individually subjected to the same minimization operation, and then produce 
geometrically distinct final patterns.  The minimized objective functions $\Phi$ and $\tilde{\Phi}$, for $D >$ 0 and  
$D =$ 0 cases respectively, indicate that these configuration sets are degenerate 
classical ground states, including disordered ones, for specific 
potential energy functions.  When $\Phi$ is used, that potential function consists of four, three and two body components, 
while for $\tilde{\Phi}$ only two-body components arise.  This offers a specific constructive method to achieve 
degenerate-ground-state potentials, a subject recently discussed by S\"{u}t\'{o}~\cite{Su05}.}

\par{ The present collective variable approach may also supply some insight into the existence of potentials whose 
classical ground states are amorphous, or at least highly irregular.  Start with a small-$\chi$ constraint case, satisfied 
with an irregular point-particle configuration.  This specific configuration will display specific $C(\boldsymbol{k})$ 
values for the unconstrained range $|\boldsymbol{k}| > K$.  Next, formally expand the constraint radius $K$ to the point 
where $\chi =$ 1, regarding the already-obtained collective variable values as constraint targets.  The irregular 
configuration in hand becomes automatically and trivially a proper solution for this extended problem.  The resulting 
potential energy specified by $\Phi$ or $\tilde{\Phi}$ then possesses that irregular configuration as its unique classical 
ground state (aside from overall translations).}

\section*{Acknowledgements}

\par{  The authors thank Andrea Gabrielli and Michael Joyce for discussions concerning several aspects of the work 
reported in this paper.  We gratefully acknowledge D. K. Stillinger for carrying out preliminary calculations that have 
provided useful guidelines for the work reported in this paper.  S.T. and F.H.S. gratefully acknowledge the support of 
the Office of Basic Energy Sciences, DOE, under Grant No. DE-FG02-04ER46108.  O.U.U. gratefully acknowledges the support 
of the Department of Energy CSGF fellowship.}

\end{document}